\documentclass[pra,twocolumn, showpacs,showkeys,secnumarabic,amssymb,amsmath,superscriptaddress]{revtex4-1}
\usepackage{amssymb,amsmath}
\usepackage{color}
\usepackage{graphics}
\usepackage{graphicx}
\usepackage{enumitem}
\usepackage{bm}
\begin{document}

\title{Time dynamics of quantum coherence and monogamy in a non-Markovian environment}

\author{Chandrashekar Radhakrishnan}
\email{chandrashekar.radhakrishnan@nyu.edu}
\affiliation{New York University Shanghai, 1555 Century Ave, Pudong, Shanghai, 200122, China.}
\affiliation{NYU-ECNU Institute of Physics at NYU Shanghai, 3663 Zhongshan Road North, Shanghai, 200062, China.}
\author{Po-Wen Chen}
\email{powen@iner.gov.tw}
\affiliation{Physics Division, Institute of Nuclear Energy Research, Longtan, Taoyuan 32546, Taiwan.}
\author{Segar Jambulingam}
\email{segar@rkmvc.ac.in}
\affiliation{New York University Shanghai, 1555 Century Ave, Pudong, Shanghai, 200122, China.}
\affiliation{Department of Physics, Ramakrishna Mission Vivekananda College, Mylapore, Chennai, 600004, India.}
\author{Tim Byrnes}
\email{tim.byrnes@nyu.edu}
\affiliation{State Key Laboratory of Precision Spectroscopy, School of Physical and Material Sciences, 
East China Normal University, Shanghai, 200062, China.}
\affiliation{New York University Shanghai, 1555 Century Ave, Pudong, Shanghai, 200122, China.}
\affiliation{NYU-ECNU Institute of Physics at NYU Shanghai, 3663 Zhongshan Road North, Shanghai, 200062, China.}
\affiliation{National Institute of Informatics, 2-1-2 Hitotsubashi, Chiyoda-ku, Tokyo, 101-8430, Japan.}
\affiliation{Department of Physics, New York University, New York, NY, 10003, USA.}
\author{Md.~Manirul Ali}
\email{mani@cts.nthu.edu.tw : Corresponding author}
\affiliation{Physics Division, National Center for Theoretical Sciences, National Tsing Hua University,
Hsinchu 30013, Taiwan.}

\begin{abstract}
The time evolution of the distribution and shareability of quantum coherence of a tripartite system in a 
non-Markovian environment is examined.  The total coherence can be decomposed into various contributions, ranging from local, global bipartite and global tripartite, which characterize the type of state.  We identify coherence revivals for non-Markovian systems for all the contributions of coherence.  The local coherence is found to be much more robust under the environmental
coupling due to an effective smaller coupling to the reservoir.  This allows us to devise a characterization of a quantum state in terms of a coherence tuple on a multipartite state simply by examining various combinations of reservoir couplings. 
The effect of the environment on the shareability of quantum coherence, as defined
using the monogamy of coherence, is investigated and found that the sign of the monogamy is a preserved quantity under the decoherence.  
We conjecture that the monogamy of coherence is a conserved property under local incoherent processes.  
\end{abstract}

\pacs{0.365.Yz, 0.367.-a}

\maketitle

%
%
%

\section{Introduction}
 
Coherence has been a central concept in quantum physics since the introduction of wave-particle duality.  For many 
years the study of quantum coherence was investigated in the context of phase space 
distributions \cite{glauber1963coherent,sudarshan1963equivalence} and higher order correlation functions
\cite{scully1999quantum}.  Recently coherence was quantified 
in a rigorous sense by Baumgratz, Cramer, and Plenio \cite{baumgratz2014quantifying}, and improved upon through several works 
\cite{yadin2016quantum,winter2016operational,chitambar2016critical,du2015conditions,cheng2015complementarity}. 
In the context of these works, coherence is now viewed as a quantum characteristic alongside other quantities 
such as discord, entanglement, steerability, and non-local correlations \cite{adesso2016measures}. It has been 
investigated in a variety of different systems such as Bose-Einstein condensates \cite{opanchuk2016quantifying}, 
cavity optomechanics \cite{zheng2016detecting,man2015cavity}, and spin systems 
\cite{karpat2014quantum,malvezzi2016quantum,radhakrishnan2017quantum,radhakrishnan2017quantum2}. 

Coherence, alongside many of the other quantum properties, are often studied without explicitly 
specifying the effect of the external environment on the system.  Under experimentally realistic situations, 
the environment will cause a time varying evolution towards a mixed state \cite{breuer2002theory}.  
A system can exhibit 
Markovian or non-Markovian dynamics depending on whether it is weakly or strongly coupled to the 
environment. 
Entanglement was the first type of quantum correlation whose dynamics was explored in this context 
\cite{maniscalco2006non,bellomo2007non,bellomo2008entanglement,li2010entanglement}. 
Several studies have shown \cite{almeida2007environment,lopez2008sudden,yu2009sudden,mazzola2009sudden} 
that the entanglement of a quantum system in a Markovian environment experiences 
an exponential decay with time. 
In a non-Markovian environment, the entanglement may reappear after a time period 
of complete disappearance 
\cite{bellomo2007non,mazzola2009sudden}, a feature referred to as entanglement revival.
Later several studies investigated the dynamics of many other quantum correlations 
\cite{fanchini2010non,man2011quantum,franco2012revival,franco2013dynamics,chanda2016delineating}.  

Entanglement and discord are purely inter-particle in nature, hence require at least two subsystems, 
but quantum coherence has 
the unique property \cite{radhakrishnan2016distribution,radhakrishnan2016coherence} that it can 
exist both at the 
inter-particle and intra-particle levels. 
The ``intrinsic'' or ``global'' coherence arises is inter-particle in nature and happens due 
to superposition between two qubits.  Meanwhile the superposition of the quantum levels 
within a single subsystem results in the local coherence.  
These two forms of coherence have a complementary nature and cannot exceed the total coherence in the system \cite{radhakrishnan2016distribution,radhakrishnan2016coherence,tan2016unified}. Extensions of this idea can be made to multiparticle systems, in particular for a tripartite system the coherence between two subsystems limits the amount of coherence 
the third subsystem can share with these systems. The monogamy of coherence then measures the shareability of coherence, and is the difference between the pair-wise and the multipartite intrinsic coherences \cite{radhakrishnan2016distribution,radhakrishnan2016coherence}. 

In this paper, we investigate the time dynamics of the distribution of coherence under a non-Markovian environment.  Our 
aim will be to examine the response of the various types of coherence, since as local and global, and see how susceptible they 
are under the incoherent operations induced by a reservoir.  By changing our parameters we can equally study the Markovian limit of the environment, which will 
allow us to study a variety of different scenarios.   We furthermore investigate the dynamics when the reservoir only partially couples to the state, on particular sites.  This leads us to devise a method for understanding the nature of the coherence simply by examining the response of the total coherence by adding successive environmental couplings to the whole system.  We also study the shareability of coherence by using the monogamy of coherence.  This is another characteristic that occurs only for systems with at least three particles, and is a identifier of the type of correlations that are present in the system.

\section{Description of the model}  

\begin{figure}[t]
\includegraphics[width=\linewidth]{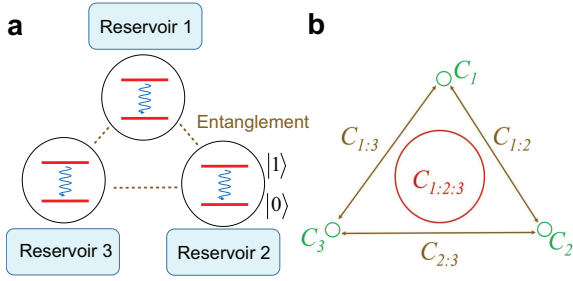} 
\caption{Coherence in a tripartite system.  (a) The model examined in this paper.  Three qubits are individually coupled to dissipative reservoirs which creates a decay from the $ | 1 \rangle $ to the $ | 0 \rangle $ state.  The reservoirs can be coupled in a Markovian or non-Markovian way, depending upon the parameters of Hamiltonian (\ref{qrhamiltonian}).  The initial state is generally considered to be a tripartite entangled state, of the form of a $ W $, $ GHZ $, or $ W \bar{W}  $ state.  (b) The coherence distribution in a tripartite system.  The local coherence $ C_j $, the bipartite global coherences $ C_{j:k} $, and the tripartite global coherence $ C_{1:2:3} $ are as marked.  }  
\label{fig1}
\end{figure}

We consider a system of three non-interacting parts, each consisting of 
a qubit interacting with a local bosonic reservoir (see Fig. \ref{fig1}(a)). The Hamiltonian of the qubits and reservoir reads
\begin{equation}
H = \sum_{j=1}^{3} \bigg[ \omega_{0}^{j} \sigma_{+}^{j} \sigma_{-}^{j} + \sum_{k} \omega_{k} b_{jk}^{\dag} b_{jk} 
      + (\sigma_{+}^{j} B_{j} + \sigma_{-}^{j} B^{\dag}_{j}) \bigg]
\label{qrhamiltonian}     
\end{equation}
with $B_{j}=\sum_{k}g_{k} b_{jk}$,
where $\sigma_{\pm}^{j}$ are the raising and lowering operators of the two level atom and 
$\omega_{0}$ is the transition frequency of the two level system.  The index $k$ labels the 
field modes of the reservoir with frequencies $\omega_{k}$, $b^{\dag}_{jk}$ ($b_{jk}$) is 
the creation (annihilation) operator for the reservoir for the $ j $th qubit, and $g_{k}$ is the coupling strength between the qubit 
and the $k$th mode of the environment. 
This model can be solved exactly at zero-temperature \cite{garraway1997decay}. 
The dynamics of each non-interacting part can be represented by the reduced density matrix
\begin{equation}
  \rho(t)=
  \left( {\begin{array}{cc}
  1-\rho_{11} |h(t)|^{2} & \rho_{01}  h(t)\\
   \rho_{10}  h^{*}(t) & \rho_{11} |h(t)|^{2}\\
  \end{array} } \right),
\label{dynamicsmatrix}
\end{equation}
where $h(t)$ is the time evolution given by 
\begin{equation*}
\frac{{\rm d} h(t)}{{\rm d} t} = - \int_{t_{0}}^{t} {\rm d} \tau f(t - \tau) h(\tau).
\end{equation*}
The correlation function $f(t - \tau)$ is related to the spectral density $J(\omega)$ of the reservoir as 
$f(t-\tau) = \int {\rm d} \omega  J(\omega) e^{i(\omega_{0} - \omega)(t - \tau)}$.  
In this work we consider a Lorentzian spectral density 
\begin{equation}
J(\omega) =  \frac{1}{2 \pi} \frac{\gamma_{0} \lambda^{2}}{(\omega_{0} - \omega - \Delta)^{2} + \lambda^{2}}.
\label{spectraldensity}
\end{equation}
for which the single qubit evolution $h(t)$ is well known \cite{breuer2002theory,maniscalco2006non,li2010entanglement} 
\begin{equation}
h(t) = e^{-(\lambda - i \Delta) t} \left[ \cosh \left(\frac{\Omega t}{2}\right) 
+ \frac{\lambda - i \Delta}{\Omega} 
\sinh \left( \frac{\Omega t}{2} \right) \right]^{2}, 
\label{evolutionequation}			
\end{equation}
where $\Omega = \sqrt{(\lambda - i \Delta)^{2}-2 \gamma_{0} \lambda}$. 
The spectral width of the reservoir $\lambda$ characterizes the reservoir correlation 
time via the relation $\tau_{1} =  \lambda^{-1}$.
Meanwhile the microscopic system-reservoir coupling $\gamma_{0}$ is the 
inverse of the relaxation time $\tau_{2}$.
The Markovian and the non-Markovian regimes can be identified from the 
relationship between these time scales.  
When $\gamma_{0} < \lambda/2$ ($\tau_{2} > 2 \tau_{1}$), the system is weakly 
coupled to the reservoir and the dynamics 
is Markovian. The non-Markovian effects due to the strong coupling regime 
arises when 
$\gamma_{0} > \lambda/2$ ($\tau_{2} < 2 \tau_{1}$).

\section{Local and global coherence}

We investigate a non-interacting three qubit system coupled to individual bosonic reservoirs as described in the previous section.  The three qubits are initialized in various states and the subsequent time evolution is examined.  To 
characterize the different types of coherence, we use the relative entropy \cite{baumgratz2014quantifying}.  The total coherence in the system is given by
\begin{equation}
C(\rho) =  \min_{\sigma \in \mathcal{I}} S(\rho \| \sigma) = S(\rho_{d}) - S(\rho),
\label{totalcoherence}
\end{equation}
where $\mathcal{I}$ is the set of incoherent state and $\rho_{d}$ is the diagonal matrix of the density
 matrix. Here $\rho_{d}$ is the diagonal matrix of $\rho$ in the basis $|0\rangle$ and $|1 \rangle$. 
It is only logical to investigate the process in the $\sigma^{z}$-basis since the dynamics is entirely 
described in that basis as we notice through Eqn. \ref{dynamicsmatrix} 
The local coherence is then found using the relation \cite{tan2016unified}
\begin{equation}
C_{L} = S(\pi(\rho) \| [\pi(\rho)]^{d}).
\label{localcoherence}
\end{equation}
where $\pi(\rho) = \rho_{1} \otimes \rho_{2} \otimes \rho_{3}$.   
From (\ref{totalcoherence}) and the (\ref{localcoherence}) one can find the global coherence by simply taking the difference
\begin{align}
C_{G} = C - C_{L}.
\end{align}
In a tripartite system the global coherence can be further decomposed in to three-way and two-way global 
coherences. The expression for these coherences are 
\begin{eqnarray}
C_{TG} &\equiv& C_{1:2:3} = C_{2:3} + C_{1:23}, \\
C_{BG} &=& C_{1:2} + C_{1:3} + C_{2:3}, \\
C_{i:j} &=& C_{G} (\rho_{ij}).
\end{eqnarray}
Here $C_{1:23}$ is the intrinsic coherence
between the qubit $\rho_1$ and the bipartite system $\rho_{23}$. If the loss of any one of the qubits completely decoheres 
the system then it is said to have a three-way or purely tripartite global coherence $(C_{TG})$. Conversely if the loss 
of any two qubits causes complete decoherence then we have a two way or bipartite global coherence $(C_{BG})$.

\begin{figure}[t]
\includegraphics[width=\linewidth]{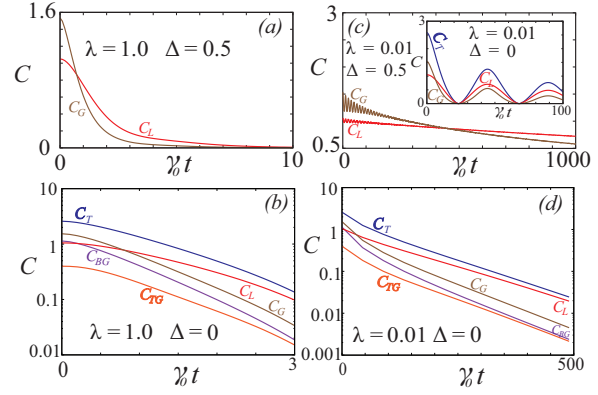} 
\caption{Time dynamics of coherence for the $ W \bar{W} $ state.  (a) The local coherence $ C_L $ and global coherence $ C_G $ in the Markovian regime. Parameters
are $ \lambda = 1, \Delta = 0.5 $.  (b) Semi-logarithmic plot for the same parameters as (a), including additionally the total coherence $ C_T $, bipartite coherence $ C_{BG} $, tripartite coherence $ C_{TG} $.  (c) Various coherences as marked in the non-Markovian 
regime.  Parameters are $ \lambda = 0.01$ and $ \Delta = 0.5 $ (main figure), $ \Delta = 0 $ (inset).  (d) Semi-logarithmic plot for the non-Markovian regime for the same parameters as the main plot of (c) for the coherences marked.}  
\label{fig2}
\end{figure}

For the initial state we use the $ W \bar{W}  $-state defined as 
\begin{align}
| W \bar{W}  \rangle = \frac{|W \rangle + |\bar{W} \rangle}{\sqrt{2}}
\end{align}
where
\begin{align}
|W \rangle &  = [|001 \rangle + |010 \rangle + |100 \rangle]/\sqrt{3} \nonumber \\
| \bar{W} \rangle & = [|011 \rangle + |101 \rangle + |110 \rangle]/\sqrt{3} .
\end{align}
This is particularly interesting as it has all types of different coherences, both local and global, distributed in both
a tripartite and bipartite manner.  This is in contrast to either $GHZ$ or $W$ states, which have zero local coherence, and 
are purely tripartite and bipartite entangled \cite{radhakrishnan2016distribution}.  By examining a state with all types of
coherences this gives a convenient way of examining the time dynamics of the various contributions.  

The dynamics of the $W \bar{W}$ state in the 
Markovian regime is illustrated in Fig. \ref{fig2}(a). We see that the local and global
coherence exhibits exponential decay but at different rates.  To find the decay rate we plot the coherence on a semi-logarithmic plot as a function of the dimensionless time $\gamma_{0} t$ as shown in  Fig \ref{fig2}(b).  The gradient of the curve on the semi-logarithmic plot gives the decay rate at a particular time.  We find that at any given time
the local coherence has a lower decay rate than the global coherence.  This is the expected result since by its very nature, 
local coherence is only present at each qubit site, which is coupled locally to the reservoir.  Thus the local coherence only experiences effectively one reservoir at a time, whereas the global coherence is distributed across the whole system.  It is thus affected by all reservoirs at the same time, and should therefore decay at a faster rate.  

We now consider the strongly non-Markovian limit, which gives richer dynamics to all types of coherence.  
The non-Markovian dynamics of the $W \bar{W}$ state both with detuning and without detuning is 
given in Fig. \ref{fig2}(c).  From the plots we 
observe that both the local and global coherence have different dynamical behavior. In particular we see the decay and 
revival of the coherence in an analogous way to that observed with entanglement \cite{bellomo2007non,mazzola2009sudden}.  
We can accordingly call this phenomena ``coherence revival'', since coherence spontaneously re-enters the system after being
initially destroyed by the bath.  For the case with zero detuning the coherence in fact is completely destroyed in all forms, and then 
spontaneously reappears.  The general phenomenology of the coherence in the non-Markovian regime is that it oscillates 
with a decaying envelope.  Thus the quantum coherence which is oscillatory at shorter time scale has an exponential 
decay at the longer time scale.  To find the decay rate we trace this exponential envelope by performing a 
linear fit of the logarithm of the coherence at the maximum point as a function time.  The slope of the semi-logarithmic 
plots in Fig. \ref{fig2}(d) then give the decay rate for the non-Markovian case.  Here too we find that the local coherence 
has a slower decay compared with the global coherence, due to only a single reservoir acting on the local coherence, 
in comparison to multiple reservoirs acting on global coherence.  

In Figs. \ref{fig2}(b)(d) we compare the decay rate of the local coherence with the total bipartite 
global coherence $C_{BG}$ and the tripartite global coherence $C_{TG}$. 
From the linear fit we observe that the local coherence has the lowest decay rate.
Further we notice that the total coherence has a lower decay rate compared with any one of its 
individual components.  Also the total global coherence has a lower decay rate compared with the 
bipartite and the tripartite global coherence.  Both these observations are along the expected lines 
since the individual components have a faster decay rate than their combined value.

\begin{figure}[t]
\includegraphics[width=\linewidth]{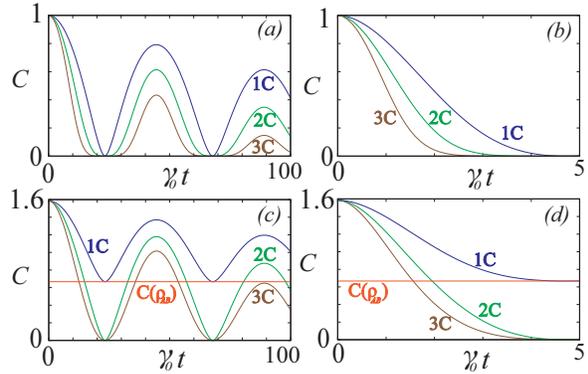}
\caption{Time dynamics of coherence for the $ GHZ $ and $ W $ states, for various reservoir couplings.  
(a)Non-Markovian evolution for the  $GHZ$ state for the number of reservoir couplings as marked.  Parameters used are 
$\lambda = 0.01$ and $\Delta =0$. (b) Markovian evolution with parameters used $\lambda = 1.0$ and $\Delta =0$.
(c) Non-Markovian evolution for the  $W$ state.  Parameters used are $\lambda = 0.01$ and $\Delta =0$. 
(d) Markovian evolution with parameters used $\lambda = 1.0$ and $\Delta =0$. }  
\label{fig3}
\end{figure}

\section{Partial coupling to the reservoir}

In the investigations so far described, all three qubits were connected 
to an external reservoir. Now 
we would like to examine the situation where only some of the qubits are connected to the reservoir. 
This can be achieved by changing the decay time of the couplings on the different qubits.  For example, 
the single channel decay regime can be defined as when the decay time of qubit 1 is much less than the remaining qubits
and also the observation time scale  $\tau$ 
\begin{align}
t_{1} < \tau \ll t_{2,3}
\end{align}
where  $t_j$ is the decay time of the $ j$th qubit. 

The states that we will examine here are the $W$ and the $GHZ$ states, defined as
\begin{align}
|GHZ \rangle = [|000 \rangle + |111 \rangle]/\sqrt{2} .
\end{align}
These are known to have a different structure of entanglement, and therefore its coherence properties can be expected to 
be different.  The $GHZ$
state is considered to be a genuinely tripartite entangled system, whereas the $W$ state is bipartite entangled, and 
are unrelated 
under local operations and classical communications.  
By changing the local couplings to the reservoir, it is reasonable to expect that these respond differently given 
the considerations of the previous section.   

In Figs. \ref{fig3}(a)(b) we show the variation of quantum coherence for the 
one, two, and three channels for 
both the Markovian and the non-Markovian regimes with an initial $GHZ$ state.  
We find that under the Markovian approximation the quantum coherence always vanishes to zero in the long-time limit.  
For the non-Markovian case we observe that 
the quantum coherence oscillates with time on the shorter time scale, while decaying in the long-time limit.  
The single channel coherence oscillations decays slower and rises faster 
in comparison with the two and the three channel cases.  In the three channel case the 
coherence falls to zero and remains so for a particular length 
of time. This behavior is because in the single channel only one qubit is 
directly in contact with the environment and the 
other two qubits are influenced by the environment due to their coherent connection with the first one.

The behavior of $W$ states is shown in Figs. \ref{fig3}(c)(d) for the Markovian and 
the non-Markovian cases respectively.  From the 
plots we observe that the three and the two channel coherence goes to zero but the 
single channel case attains a 
steady state value of $C=2/3$ for the Markovian case.  
In the non-Markovian case the quantum coherence oscillates,
but the oscillatory minimum is zero for the two and three channel cases but 
for the single channel case
it is equal to $C=2/3$.  
This is because the coherence in the $W$ state is distributed in a bipartite manner and 
the environment is acting on a single qubit in the single channel case.  In the long-time limit the $ W $ state evolves to the mixed state
\begin{align}
\rho_{\text{1C}} & ( t \rightarrow \infty)  = \sum_i M_i | W \rangle \langle W | M^\dagger_i \nonumber \\
= &  |0 \rangle \langle 0 | \otimes \Big[ \frac{2}{3}
\left( \frac{| 01 \rangle + | 10 \rangle}{\sqrt{2}} \right) \left( \frac{\langle 01 | + \langle 10 |}{\sqrt{2}} \right) \nonumber \\
& + \frac{1}{3} | 00 \rangle \langle 00 | \Big]
\label{onecrho}
\end{align}
where we have taken the measurement operators on the first qubit
\begin{align}
M_0 & = | 0 \rangle \langle 0 |  \nonumber \\
M_1 & = | 0 \rangle \langle 1 |
\end{align}
corresponding to a decay of a qubit.  As we can see from (\ref{onecrho}) the decoherence of the single qubit does not 
completely destroy the total coherence in the system.  
The value $C=2/3$ attained in the steady state limit for the 
Markovian situation and as the oscillatory minimum in the non-Markovian 
situation is the coherence between the two qubits
which are not influenced by the environment in any way.  The value $C=2/3$ correspond to the 
probability of the Bell state in (\ref{onecrho}), which is the only contribution to the coherence in this case. 
For the two and three channel coherences, it is easy to see that further measurement of (\ref{onecrho})
will completely collapse the Bell state superposition, hence eventually there is zero coherence in the long-time 
limit.  We note that this is in stark contrast to the coherence in a $GHZ$ 
state which is distributed in a completely tripartite manner such that the 
decoherence of a single qubit will always destroy the 
total coherence in the system.

From the above results we observe that by coupling the reservoir in different
ways, information can be obtained about how
coherence is distributed in a multipartite system.  For tripartite 
systems the quantum coherence can be characterized in the manner shown in Fig. \ref{fig1}(b).  
Firstly, there are three local coherences, one for each qubit.  
Next, there are the three bipartite 
global coherences $C_{1:2}, C_{1:3}$ and $C_{2:3}$, according to the 
pairings of each qubit.  Lastly, there is a genuinely tripartite 
global coherence $C_{1:2:3}$.  The
total coherence can be distributed in only these seven different contributions.  
Hence we can construct a seven-tuple 
\begin{align}
{\cal C} = \{C_{1}, C_{2}, C_{3}, C_{1:2}, C_{1:3}, C_{2:3}, C_{1:2:3} \} 
\label{seventuple}
\end{align}
which 
contains all the information about the 
distribution of coherence in the system. Clearly the above procedure can be generalized in an analogous
way for a multipartite system.   If we do not have a knowledge 
of an initial quantum state, we can 
reverse engineer the state if we have the time evolution dynamics of the 
quantum coherence which can provide 
us with this seven-tuple.  For example, by coupling a reservoir to qubit 1 and looking at the steady state, the 
coherence with all terms involving qubit 1 will be destroyed, yielding
\begin{align}
{\cal C}\left|_{\Pi_1} \right.= \{0, C_{2}, C_{3},0, 0, C_{2:3}, 0 \} ,
\end{align}
where $ \Pi_j $ denotes a measurement on the $ j $th qubit. 
Similarly the coherence involving qubit 2 can be destroyed by applying two reservoirs
\begin{align}
{\cal C}\left|_{\Pi_1 \Pi_2} \right. = \{0,0, C_{3},0, 0,0, 0 \} ,
\end{align}
which yields the local coherence on qubit 3 alone.  By combining all possible measurement combinations $ I, \Pi_1, \Pi_2, \Pi_3, \Pi_1 \Pi_2, \Pi_1 \Pi_3,  \Pi_2 \Pi_3 $ we can deduce all the coherences within the system.  The number of possible measurement combinations 
is always guaranteed to be the same as the number of different coherences because the coherences appear as all $n$-way groupings of the subsystems, which is the same as for the measurements.  This allows a consistent evaluation of coherence in any multipartite system.

\begin{figure}[t]
\includegraphics[width=\linewidth]{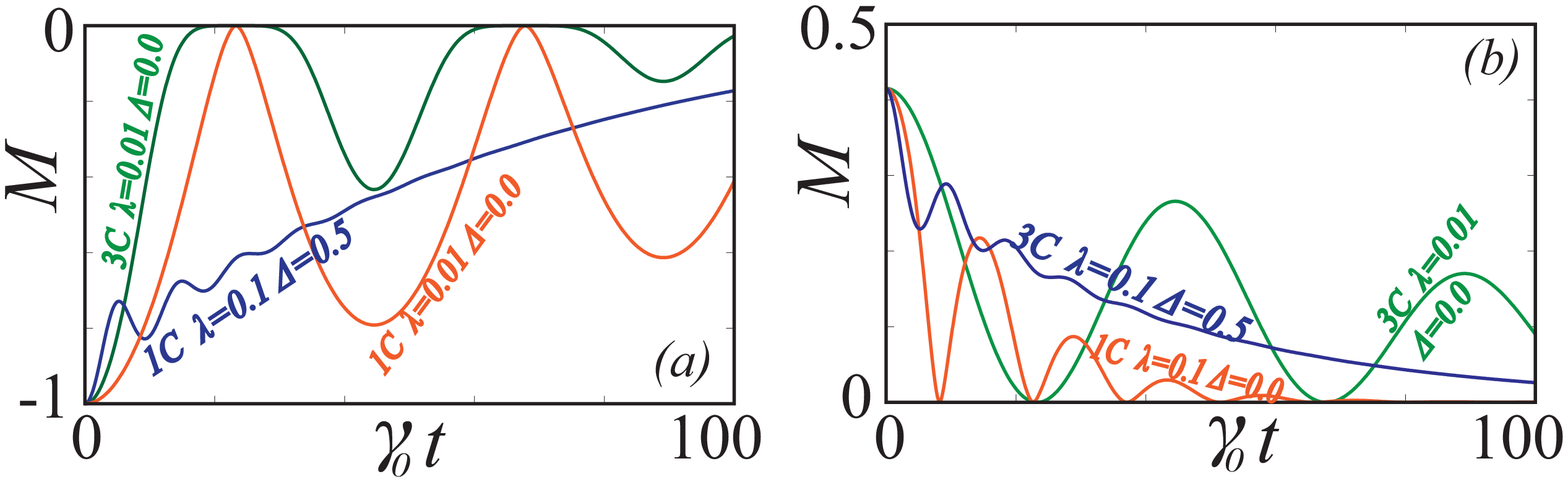}
\caption{The time evolution of the monogamy of coherence for (a) $ GHZ$ and (b) $W $ states under various conditions as labeled. }  
\label{fig4}
\end{figure}

\section{Monogamy of coherence}

We have seen that in a multipartite system the global coherence can be 
further decomposed into the bipartite contribution, 
tripartite contribution and so on up to the $N$-partite contributions \cite{radhakrishnan2016distribution,tan2016unified}.  
Quantum systems thus have a unique way sharing
coherence which is captured by the monogamy of coherence introduced in Ref. \cite{radhakrishnan2016distribution}, 
in analogy with the monogamy of entanglement \cite{coffman2000distributed,koashi2004monogamy}.  For a 
tripartite system the monogamy of coherence reads:
\begin{equation}
M = C_{1:2} + C_{1:3} - C_{1:23}
\end{equation}
Here $C_{1:2}$ ($C_{1:3}$) denotes the global coherence between the qubits 
1 \& 2 (1 \& 3) and $C_{1:23}$ is the global coherence 
between qubit 1 and the bipartite block 23. In a genuinely tripartite coherent system, the system is described as being monogamous and we observe $M\leq 0$.  When $M>0$, the 
bipartite coherence is more dominant and the system is polygamous.
  $ GHZ $ and $W $  states are archetypal examples of a polygamous and monogamous state respectively. 

In Fig. \ref{fig4} we calculate the time evolution of the monogamy of coherence of the $GHZ$ and $W$ states 
under various non-Markovian conditions.  In all cases that we have calculated we observe that the 
monogamy of coherence does not change sign, and retains its initial character.  In the strongly non-Markovian regime, 
the monogamy of coherence can become zero, particularly at points where the overall coherence, and hence its constituents vanish
\begin{align}
C_{1:2} = C_{1:3} = C_{1:23} = 0.
\end{align}
We 
have verified that there is no violation of the sign preservation by examining points where $ M $ is small and 
have not found any exceptions.   This can be understood to be due to the fact that the quantum symmetries 
of the system regarding the spatial distribution do not change under time evolution.  The form of the reservoir
coupling is strictly in a local fashion, and in a general sense corresponds to a local operation.  Since the 
$GHZ$ and $W$ states are known to be either polygamous and monogamous under local operations  \cite{radhakrishnan2016distribution}, and our reservoir model 
falls under the same category of operations.  This means that $ M $ correctly characterizes the 
polygamous or monogamous nature of the state.  We thus conjecture that a quantum state under a local time evolution 
process will preserve the monogamy of coherence in a multipartite system.

\section{Summary and conclusions}

The time evolution of quantum coherence of a three qubit system each interacting 
with a local environment was been investigated in 
both the Markovian and non-Markovian limits.
For the tripartite $W \bar{W}$ system the total coherence can be decomposed into the 
local and the global components, according to whether the coherence are intra- or inter-qubit in nature. 
Due to the tripartite nature of the system, the global 
coherence can be decomposed into the bipartite global coherence and the tripartite 
global coherence.  In analogy to entanglement revivals, we observed coherence revivals in the non-Markovian case, where the coherence can return to the system from the reservoir.  In several cases this was observed to occur even after the coherence collapsed to zero.   The general observation from this is that the local coherence decays much slower in comparison
with other forms of coherence.  This behavior is irrespective of whether the 
dynamics is Markovian or non-Markovian. This points to the fact that local coherence is much more
robust in the presence of decoherence than global coherence. Previous studies 
\cite{simon2002robustness,borras2009robustness,ali2014robustness} have shown that the 
robustness varies with the nature of a quantum state.  Contrast to these works we show that the 
different contributions of coherence decay at different rates in the same quantum state. Therefore, localizing the coherence can be one effective strategy towards extending 
the life-time of a quantum state in physical systems.  By temporarily storing it in this form, and converting it to global coherence
according to the complementary nature of coherence, this can be an effective strategy towards preserving coherence in the system. 
It is interesting to note that recently experiments have been carried out in which the interconversion of quantum coherence 
in to other quantum correlations like discord and entanglement have been the main focus of the study.  Particularly in \cite{qiao2017activation,wu2017experimental}
the local coherence in a quantum system has been converted into discord which was again successfully 
steered into local coherence. This experiment establishes the feasibility of interconversion of local and global coherence which may 
help us to prolong the quantum correlations in a system.

We also investigated the response of the system to changing the number of reservoir couplings throughout the system. For both the $GHZ$ and $W$ states, the three channel
coherence decayed the state more rapidly in comparison with
the single channel coherence.  By changing the number of reservoir couplings to the system,
it was found that various types of coherence could be selectively destroyed, giving a characterization of the 
coherence in the system.  For example, with only one reservoir coupled to the system, 
the coherence does not go to zero for $W$ state but 
it does so for a $GHZ$ state.  This is due to the way in which 
coherence is distributed among the qubits.  This leads us to the characterization of the coherence in a 
multipartite system according to (\ref{seventuple}).  By coupling reservoirs in various configurations one can selectively ``turn off''
the coherence for various contributions.  Since the number of ways of reservoir couplings is always guaranteed to be the same as the 
number of elements in the coherence tuple, this allows a powerful way of characterizing the coherence in a multipartite 
system.  Finally from the time evolution of the monogamy of coherence we found that 
the system preserves it initial nature of either monogamy 
or polygamy.  This can be explained due to the local couplings of the reservoirs, which can be viewed as incoherent local operations on the system.  Such operations are known not to change the character of the system in term of monogamy or polygamy in the context of entanglement.  We find that this is true also in the coherence case, and conjecture that the sign of the monogamy of coherence is a preserved quantity for local operations.  We found no numerical violations to this, for all the parameters and states that were tried.  

The extension of coherence to non-unitary evolution has shown that we can obtain several interesting characterizations of the original quantum state.  The rate of decay of the various coherences gives the robustness of the state under environmental influence.  By examining its distribution one can directly observe that the state evolves in such a way that certain components of its coherence decay faster than others.  Thus under partial decoherence one might expect to find that the more robust types of coherence are predominantly left.  By looking in the long-time limit one can even completely characterize the distribution of the coherence.  Remarkably, the number of measurement combinations is equal to the number of coherences, which show that this is always possible using this prescription.  One possible extension of our work is to look at more complex systems, which can be used as a method of characterizing many-body quantum states.  This program has already been started in several works such as Refs. \cite{radhakrishnan2017quantum2,radhakrishnan2017quantum,karpat2014quantum,malvezzi2016quantum} and 
could be used in contexts such as detecting quantum phase transitions.

\section{Acknowledgements} 
MMA acknowledges the support from the Ministry of Science and Technology of Taiwan 
and the Physics Division of National Center for Theoretical Sciences, Taiwan. 
PWC would like to acknowledge support from the Excellent Research Projects of 
Division of Physics, Institute of Nuclear Energy Research, Taiwan.
JS would like to acknowledge the receipt
of the Research and Advanced Training Fellowship-2016
by The World Academy of Sciences (TWAS) and the
NYU-ECNU Institute of Physics at NYU Shanghai for
visiting New York University Shanghai.
TB is supported by the Shanghai Research Challenge Fund; 
New York University Global Seed Grants for Collaborative Research; 
National Natural Science Foundation of China (61571301); 
the Thousand Talents Program for Distinguished Young Scholars (D1210036A); 
and the NSFC Research Fund for International Young Scientists (11650110425); 
NYU-ECNU Institute of Physics at NYU Shanghai; 
the Science and Technology Commission of Shanghai Municipality (17ZR1443600); 
and the China Science and Technology Exchange Center (NGA-16-001).


\end{document}